\documentclass[conference]{IEEEtran}
\IEEEoverridecommandlockouts
\usepackage{cite}
\usepackage{amsmath,amssymb,amsfonts}
\usepackage{algorithmic}
\usepackage{graphicx}
\usepackage{subcaption}

\usepackage{textcomp}
\usepackage{xcolor}
\usepackage{hyperref} 
\usepackage{setspace}

\def\BibTeX{{\rm B\kern-.05em{\sc i\kern-.025em b}\kern-.08em
    T\kern-.1667em\lower.7ex\hbox{E}\kern-.125emX}}

\makeatletter
\def\@IEEEpubidpullup{8\baselineskip}
\makeatother

\begin{document}

\IEEEoverridecommandlockouts
\IEEEpubid{
\parbox{\columnwidth}{\vspace{-4\baselineskip}Permission to make digital or hard copies of all or part of this work for personal or classroom use is granted without fee provided that copies are not made or distributed for profit or commercial advantage and that copies bear this notice and the full citation on the first page. Copyrights for components of this work owned by others than ACM must be honored. Abstracting with credit is permitted. To copy otherwise, or republish, to post on servers or to redistribute to lists, requires prior specific permission and/or a fee. Request permissions from \href{mailto:permissions@acm.org}{permissions@acm.org}.\hfill\vspace{-0.8\baselineskip}\\
\begin{spacing}{1.2}
\small\textit{ASONAM '19}, August 27-30, 2019, Vancouver, Canada \\ 
\copyright\space2019 Association for Computing Machinery. \\
ACM ISBN 978-1-4503-6868-1/19/08 \\
\url{http://dx.doi.org/10.1145/3341161.3342870} 
\end{spacing}
\hfill}
\hspace{0.9\columnsep}\makebox[\columnwidth]{\hfill}}
\IEEEpubidadjcol

\title{A Large-Scale Empirical Study of Geotagging Behavior on Twitter\\
}

\author{\IEEEauthorblockN{1\textsuperscript{st} Binxuan Huang}
\IEEEauthorblockA{\textit{School of Computer Science} \\
\textit{Carnegie Mellon University }\\
Pittsburgh, United States \\
binxuanh@cs.cmu.edu}
\and
\IEEEauthorblockN{2\textsuperscript{nd} Kathleen M. Carley}
\IEEEauthorblockA{\textit{School of Computer Science} \\
\textit{Carnegie Mellon University}\\
Pittsburgh, United States s \\
kathleen.carley@cs.cmu.edu}
}

\maketitle

\begin{abstract}
Geotagging on social media has become an important proxy for understanding people's mobility and social events. Research that uses geotags to infer public opinions relies on several key assumptions about the behavior of geotagged and non-geotagged users. However, these assumptions have not been fully validated. Lack of understanding the geotagging behavior prohibits people further utilizing it. In this paper, we present an empirical study of geotagging behavior on Twitter based on more than 40 billion tweets collected from 20 million users. There are three main findings that may challenge these common assumptions. Firstly, different groups of users have different geotagging preferences. For example, less than 3\% of users speaking in Korean are geotagged, while more than 40\% of users speaking in Indonesian use geotags. Secondly, users who report their locations in profiles are more likely to use geotags, which may affects the generability of those location prediction systems on non-geotagged users. Thirdly, strong homophily effect exists in users' geotagging behavior, that users tend to connect to friends with similar geotagging preferences.
\end{abstract}


\section{Introduction}
The increasing volume of user generated content on social media provides a great opportunity to understand the underlying user behavior. As a popular way of sharing users' current statuses, geotagging is widely used on many social networking sites like Facebook, Foursquare and Brightkite \cite{li2009analysis, noulas2011empirical}. These social networking sites allow users to tag location information to their posts so that users' friends can see where they are. 

Analyzing geotagged posts helps us understand human behavior and social events. At the individual level, user's location information enable recommendations for relevant points of interest \cite{hu2013spatial}. At the group level, geotagging supports regional tourist attraction improvement \cite{maeda2016decision}, event detection \cite{wei2017location}, and disaster management \cite{crooks2013earthquake}. It also allows us to study human activity and movement \cite{cho2011friendship}.

Specifically on Twitter, users can provide their real-time location information by tagging places to their tweets. There are five different place types on Twitter from big to small: country, admin, city, neighborhood, and POI (place of interest). These places are predefined geo-bounding boxes in the world.  In addition, users can further choose to tag precise geo-coordinates to their tweets, which allows researchers to study human's mobility in fine granularity.


A common practice of doing research related to Twitter is to connect to Twitter's streaming API\footnote{https://developer.twitter.com/en/docs/api-reference-index} and collect data for a continuous time period. Researchers can use geographical regions to get more specific geotagged results. After obtaining tweets within these regions, researchers often use these geotagged users' opinions to infer non-geotagged local users' opinions. One simple but direct way is to use geotagged users to represent the general local public \cite{widener2014using}. The basic assumption here is that geotagged and non-geotagged users come from the same distribution, which may not be true. Different users may have different geotagging preferences. Hence, we formulate our first research question as follows:

\textbf{RQ1}: Are there any differences in terms of geotagging behavior among different users? More specifically, do people with different attributes like device, country, or language have different geotagging preferences?

Instead of using geotagged tweets to represent the regional public opinions directly, many researchers try to first predict non-geotagged users' locations. Then they could combine geotagged users with those predicted non-geotagged users to get better representatives for each region. Most of these user location prediction systems are based on learning location specific features from geotagged tweets \cite{graham2014world,huang2017predicting}. In these location classifiers, a very important feature is the self-reported location in each user's profile. Using this feature assumes that users who use geotags and who do not are equally likely to report their home locations in profiles. In our second research question, we want to examine the correlation between the geotagging behavior and the behavior of reporting locations in profiles. 

\textbf{RQ2}: Is there any correlation between the geotagging behavior and the behavior of reporting location in profile?

Researchers show individuals connected in a network are highly homophilous, that people tend to connect to friends with similar demographics like age, gender, and income \cite{mcpherson2001birds}. Users' trajectories on online social networks also present strong geographical homophily \cite{cho2011friendship}. Based on this observation, various graph-based location predictors have been developed \cite{backstrom2010find,rahimi2015twitter}. However, if non-geotagged users tend to connect to similar non-geotagged users, then it would be harder to infer their locations based on their social ties. In this last research question, we explore whether geotagging preferences are similar between friends.

\textbf{RQ3}: Is there any homophily effect between friends in terms of geotagging preference?

To answer these three research questions, we collect more than 40 billion tweets from 20 million users and conduct three levels of analyses: tweet-level, user-level, and graph-level. We will present these three level analyses after discussing related work in next section.

\section{Related Work}
\subsection{The use of geotags}
There are a lot of studies using geotagged social media for various purposes. Examples include studying human mobility via geotagged tracks \cite{jurdak2015understanding, huang2018location}, understanding regional user behavior \cite{kisilevich2010event}, and inferring user locations based on geotagged users \cite{huang2017predicting}. 

Geotagging on online social networks has been an important proxy for studying human mobility \cite{jurdak2015understanding}. Hawelka et al. presented a global study of mobility characteristics across nations based on the analysis of geotagged tweets \cite{hawelka2014geo}. Using online location check-in data and cell phone location data, Cho et al. found human movement is a combination of periodic short range movements with long range travels \cite{cho2011friendship}.

Geotagged content on social media is also a direct way to understand events and the corresponding regional users' opinions \cite{huang2018parameterized, huang2018aspect}. For example, Sakaki et al. used geotagged tweets to detect earthquakes in real-time \cite{sakaki2010earthquake}. The associated locations of earthquake related tweets are used to estimate the location of an earthquake event. Using geotagged tweets, Widener and Li found that people in low-income and low-access census tracts talk less about healthy foods with a positive sentiment on Twitter \cite{widener2014using}.

Researchers also use geotagged tweets to infer users' locations. Various algorithms have been proposed for this purpose. One common way to geolocate one Twitter user is finding location indicative words in tweets and user profile \cite{han2012geolocation, zhang2017rate}. Later, researchers developed neural network based models to learn high-level location features from tweets \cite{huang2017predicting}. Based on the location homophily of friends \cite{cho2011friendship}, graph-based location prediction systems are introduced to further enhance the performance \cite{rahimi2015twitter, qian2017probabilistic}. In most of these geolocation prediction systems, geotagged tweets are used as training data in the supervised learning procedure.

\subsection{The study of geotagging behavior}
We have discussed several examples of using geotagged social media for various purposes. However, the limited understanding of the geotagging behavior hinders us better utilizing it. Researchers have conducted several studies to examine why and how people use geotags on social media.

To better understand the underlying motivation of location tagging behavior, Kim surveyed 255 Facebook users in college and found various motivational mechanisms depending on the habitualness of their mobile phone use \cite{kim2016drives}. Similarly, Tasse et al. investigated why people geotag on Twitter and found people geotag their tweets to show off where they are or communicate with family and friends \cite{tasse2017state}.

In addition to understand why people use geotags, there are also several studies try to uncover how people use geotags on social media. Ludford et al. conducted two small-scale empirical studies to show how people share location knowledge through different location types \cite{ludford2007capturing}. Noulas et al. presented a large-scale study of user geotagging behavior on Foursquare \cite{noulas2011empirical}. They analyzed the spatio-temporal patterns of user check-in dynamics and found the general consensus of user activity at a given time and place.

Sloan and Morgan presented a closely related work \cite{sloan2015tweets}, where they analyzed the correlation between demographic characteristics and the use of geotags on Twitter. One fundamental difference between their work and ours is the data collecting strategy. They only collected a 1\% random feed from Twitter's streaming API, while we download a comprehensive set of tweets for each user as well as their following relationships. Their estimation cannot represent the true underlying distribution, since the majority of tweets for targeted users are missed in their collection. We also expand their work by considering the geotagging behavior homophily in the following network.

\section{Data}
\begin{table*}[!h]
\centering
\caption{A brief summary of our tweet collections. More than 40 billion tweets are collected from these sampled users. Only 2.31\% of these tweets are geotagged. }
\label{data_stats_table}
\begin{tabular}{|c|c|c|c|c|}\hline
\# of tweeters & \# of tweets   & \# of following ties   & \# of place-tagged tweets & \# of coordinates-tagged tweets \\ \hline
19,984,064     & 41,267,348,020 & 4,402,458,603 & 724,933,445 (1.76\%)      & 228,606,700 (0.55\%)            \\ \hline
\end{tabular}
\end{table*}
\begin{table*}[!th]
\centering
\caption{Geotagging distributions among various source platforms. Tweet sources are sorted by the number of tweets sent from them.}
\label{tweet_source}
\begin{tabular}{|l|r|r|r|}
\hline
Tweet source              & \# of non-geotagged tweets   & \# of place-tagged tweets & \# of coordinates-tagged tweets \\ \hline
Twitter for iPhone  & 15,716,820,447 (97.25\%) & 393,059,787 (2.43\%)      & 51,527,597 (0.32\%)             \\ \hline
Twitter for Android & 11,677,533,121 (97.81\%) & 219,107,978 (1.84\%)      & 42,247,513 (0.35\%)             \\ \hline
Twitter Web Client  & 4,088,127,646 (97.69\%)  & 96,643,283 (2.31\%)       & 126,639 (0.00\%)                \\ \hline
twittbot.net        & 916,067,510 (100.0\%)    & 0 (0.00\%)                   & 0 (0.00\%)                         \\ \hline
Facebook            & 769,543,040 (100.0\%)    & 0 (0.00\%)                   & 0 (0.00\%)                         \\ \hline
Twitter for iPad    & 624,738,931 (98.67\%)    & 6,979,518 (1.10\%)        & 1,420,852 (0.22\%)              \\ \hline
TweetDeck           & 526,790,924 (99.98\%)    & 25,725 (0.00\%)           & 53,311 (0.01\%)                 \\ \hline
Twitter Lite        & 500,696,593 (99.98\%)    & 64 (0.00\%)                    & 116,467 (0.02\%)                \\ \hline
Instagram           & 246,133,428 (80.89\%)   & 1,470 (0.00\%)                 & 58,140,075 (19.11\%)            \\ \hline
Others              & 5,247,435,271 (98.42\%)  & 9,115,620 (0.17\%)        & 74,974,246 (1.41\%)             \\ \hline
\end{tabular}
\end{table*}

\subsection{Data Collection}
This work uses Twitter's official API to collect 20 million users' data. We first connect to Twitter's sample streaming API\footnote{https://developer.twitter.com/en9/docs/api-reference-index} to sample real-time tweets without any filter predicates. We consider this sampled streaming data representing a random sample of the Twitter population. After we get 22,881,250 users involved in this sampled data, we query Twitter's API again and collect the timelines and following ties for these sampled users. After this step, we get 19,984,064 users whose timelines and following lists are publicly available. With such method, we can get a comprehensive dataset that containing all the information we need for sampled global users with minimal bias compared with other methods like snowball sampling \cite{biernacki1981snowball}.
Certain limitations still exist because Twitter API's regulations. First, Twitter only allows people collecting the most recent 3200 tweets for each given user. Second, even certain users' tweets appear in the sampled streaming dataset, we still cannot collect their timelines and following ties because of their privacy settings. In practise, one can also collect these tweets by sampling as Sloan and Morgan did \cite{sloan2015tweets}. However, only sampling a small proportion tweets for each user cannot guarantee a correct estimation of underlying geotagging distribution.

\subsection{Data Statistics}

As shown in Table \ref{data_stats_table}, we extract over 40 billion tweets from these 20 million Twitter users. On average, we collect 2065 tweets for each user. Because we are interested in the overall distribution of geotagged tweets among all the tweets, we have not excluded retweets in our data collection.
Among these 40 billion tweets, only 2.31\% of them are geotagged (either only contains a place tag or combined with a coordinates tag), which is higher than previous estimation \cite{graham2014world}. As expected, only a small portion (23.97\%) of these geotagged tweets have precise coordinates. In this paper, we divide geotagged tweets into two types --- place-tagged and coordinates-tagged. If a geotagged tweet has precise geo-coordinates, then we call it coordinates-tagged, otherwise place-tagged. 

Unlike tweet-level geotagging, the definition for geotagged users is not straight forward. We aggregate all the tweets for each user. If one user has posted at least one geotagged tweet, then we call this user geotagged. We also divide these geotagged users into two groups --- place-tagged users and coordinates-tagged users.  If a geotagged user has at least one coordinates-tagged tweet, then we call this user coordinates-tagged, otherwise place-tagged. 

Among these 19,984,064 users in our final dataset, 24.38\% (n=4,871,784) of them are geotagged and 12.93\% (n=2,584,042) of them are coordinates-tagged, which is much higher than previous estimation \cite{sloan2015tweets,sloan2013knowing}. This implies that given a set of users, we can at least know the locations for a quarter of them which is much more prevalent than tweet-level geotagging.

\section{Tweet-level Analysis}
Before we conduct our user-level analysis, we first look at the tweet-level geotagging distribution. We compare tweets with different attributes like tweet source and tweet language, and examine whether tweets with certain attributes would have a higher chance to be geotagged.

\subsection{Tweet Source}
There is an attribute called source in each tweet JSON object which records the utility used to post the tweet. Typical examples are ``Twitter for iPhone'', ``Twitter for Android'', and ``Twitter Web Client''. We examined the percentages of place-tagged tweets and coordinates-tagged tweets for each platform. The results are shown in Table \ref{tweet_source}. For the two most popular Twitter platforms iPhone and Android, it appears that iPhone users have a much higher probability to send geotagged tweets than Android users (2.75\% versus 2.19\%). The percentages of coordinates-tags among geotagged tweets are 13.11\% and 19.28\% for iPhone and Android respectively, which means when users choose to send geotagged tweets, Android users are more likely to associate precise coordinates than iPhone users. Because of the nature of certain Twitter sources, some of them do not allow users to share real-time geolocation information to Twitter, eg. ``twitterbot.net'' and ``Facebook''. Surprisingly, coordinates-tags widely exist in tweets from Instagram. 19.11\% of the tweets from Instagram contain coordinates-tags which is almost 35 times higher than the average percentage. Same thing also happens for Foursquare, from which 74.64\% tweets are coordinates-tagged. Though there are only 65,909,517 tweets sent from Foursquare, 49,193,479 of them are coordinates-tagged. As a result, Instagram is the tweet source with the largest number of coordinates-tagged tweets, followed by iPhone and Foursquare.

\begin{table*}[!th]
\centering
\caption{Place type distributions among top place-tagged tweet sources.}
\label{place_source}
\resizebox{\textwidth}{!}{
\begin{tabular}{|l|r|r|r|r|r|}
\hline
Tweet source              & country                                                        & admin                                                          & city                                                            & neighborhood                                                  & poi                                                          \\ \hline
Twitter for iPhone  & 6,448,752 (1.64\%)   & 47,820,014 (12.17\%) & 336,007,216 (85.49\%) & 269,817 (0.07\%)    & 2,513,988 (0.64\%) \\ \hline
Twitter for Android & 5,633,166 (2.57\%)   & 23,219,855 (10.60\%) & 188,822,442 (86.18\%) & 278,991 (0.13\%)   & 1,153,524 (0.53\%) \\ \hline
Twitter Web Client  & 14,180,029 (14.67\%) & 15,557,789 (16.10\%) & 66,777,976 (69.10\%)  & 127,489 (0.13\%)    & 0 (0.00\%)        \\ \hline
Twitter for iPad    &144,597 (2.07\%)    & 647,221 (9.27\%)     & 6,131,541 (87.85\%)   & 9,407 (0.13\%)      & 46,752 (0.67\%)  \\ \hline
Tweetbot for iOS    & 77,587 (1.40\%)    &617,288 (11.13\%)   & 3,604,772 (64.98\%)   & 1,247,505 (22.49\%) & 0 (0.00\%)           \\ \hline
Tweetbot for Mac    & 14,989 (0.95\%)     & 154,566 (9.84\%)    & 923,905 (58.79\%)     & 477,944 (30.42\%)   & 0 (0.00\%)           \\ \hline
Others              & 94,431 (2.63\%)    & 424,966 (11.82\%)   & 2,553,342 (71.01\%)   & 516,441 (14.36\%)   & 6,540 (0.18\%)    \\ \hline
\end{tabular}
}
\end{table*}

\begin{table*}[!h]
\centering
\caption{Distributions of geotags for tweets with different tweet languages (top 15)}
\label{tweet_lang}
\begin{tabular}{|l|r|r|r|}
\hline
Lang. & Non-geotagged       & Place-tagged    & Coordinates-tagged  \\ \hline
English    & 14,209,166,056 (97.04\%) & 330,133,459 (2.25\%) & 103,425,905 (0.71\%)     \\ \hline
Japanese    & 7,920,019,090 (99.36\%)  & 37,943,333 (0.48\%)  & 13,167,513 (0.17\%)      \\ \hline
Spanish    & 4,405,421,261 (97.39\%)  & 89,268,301 (1.97\%)  & 28,635,593 (0.63\%)      \\ \hline
Arabic    & 3,063,691,725 (99.29\%)  & 18,598,541 (0.60\%)  & 3,335,846 (0.11\%)       \\ \hline
Portuguese    & 2,366,206,011 (95.67\%)  & 91,288,151 (3.69\%)  & 15,848,111 (0.64\%)      \\ \hline
und   & 2,327,452,586 (97.38\%)  & 53,030,818 (2.22\%)  & 9,673,783 (0.40\%)       \\ \hline
Korean    & 1,013,674,569 (99.86\%)  & 976,321 (0.10\%)     & 436,763 (0.04\%)         \\ \hline
French    & 820,763,421 (98.05\%)    & 13,489,214 (1.61\%)  & 2,834,209 (0.34\%)       \\ \hline
Indonesian    & 778,781,264 (96.02\%)    & 16,621,968 (2.05\%)  & 15,651,480 (1.93\%)      \\ \hline
Thai    & 729,496,967 (99.00\%)    & 5,588,546 (0.76\%)   & 1,748,479 (0.24\%)       \\ \hline
Turkish    & 670,097,929 (95.38\%)    & 16,085,714 (2.29\%)  & 16,352,663 (2.33\%)      \\ \hline
Tagalog    & 473,457,971 (96.02\%)    & 16,112,019 (3.27\%) & 3,498,319 (0.71\%)       \\ \hline
Russian    & 320,550,743 (96.08\%)    & 10,132,568 (3.04\%)  & 2,929,779 (0.88\%)       \\ \hline
Italian    & 229,578,445 (97.03\%)    & 5,093,448 (2.15\%)   & 1,936,471 (0.82\%)       \\ \hline
German    & 162,995,168 (97.57\%)    & 2,847,769 (1.70\%)   & 1,214,755 (0.73\%)         \\ \hline
\end{tabular}

\end{table*}

For these place-tagged tweets, we further looked at the distribution of place type for each tweet source. There are five types of place-tags: country, admin, city, neighborhood, and poi (place of interest). We present the distributions of place-tags for different tweet sources in Table \ref{place_source}. On average, city is the most widely used place-tag among users. 83.3\% of these place-tagged tweets are associated with a city label, while only 0.34\% and 0.51\% place-tags are neighborhood and poi respectively. For tweets from Twitter web client, they are much more likely to contain tags with large regions like country or admin than tweets from other platforms. On the contrary, tweets from Tweetbot have a higher chance to be tagged with fine-grained place-tags. There is a significant association between tweet sources and types of place-tags ($\chi^2$=168241325.96, 20df, p$<$0.001).

\begin{table}[!h]
\centering
\caption{Percentages of coordinates-tagged tweets for countries (top 15).}
\label{tweet_country}
\resizebox{0.5\textwidth}{!}{
\begin{tabular}{|l|r|r|}
\hline
country & \# of geotagged & \# of coordinates-tagged \\ \hline
United States      & 320,268,573     & 61,488,648 (19.20\%)     \\ \hline
Brazil      & 117,794,897     & 19,509,860 (16.56\%)     \\ \hline
United Kingdom       & 59,983,328      & 14,585,781 (24.32\%)     \\ \hline
Japan      & 51,847,289      & 13,406,333 (25.86\%)     \\ \hline
Argentina      & 39,744,563      & 6,350,980 (15.98\%)      \\ \hline
Turkey      & 35,989,555      & 19,037,195 (52.90\%)     \\ \hline
Philippines      & 29,031,714      & 4,696,974 (16.18\%)      \\ \hline
Mexico      & 24,317,203      & 8,132,105 (33.44\%)      \\ \hline
Spain      & 22,608,661      & 6,114,047 (27.04\%)      \\ \hline
Malaysia      & 22,036,169      & 8,096,642 (36.74\%)      \\ \hline
Indonesia      & 18,581,142      & 12,357,982 (66.51\%)     \\ \hline
France      & 16,049,418      & 2,690,101 (16.76\%)     \\ \hline
Canada      & 13,142,453      & 2,987,905 (22.73\%)      \\ \hline
Russia      & 11,241,015      & 2,844,891 (25.31\%)      \\ \hline
Saudi Arabia      & 10,728,248      & 1,568,910 (14.62\%)      \\ \hline
\end{tabular}
}
\end{table}

\subsection{Tweet Country}
Each geotagged tweet is assigned a country label depending on which country it belongs to. We conduct similar analysis of geotagging distributions for different countries. The only difference is that we do not have the percentages of non-geotagged tweets for these countries because of lack of country labels. Table \ref{tweet_country} shows the percentages of coordinates-tagged tweets for different countries ($\chi^2$=50145593.50, 14df, p$<$0.001). For most of these countries, the percentages of coordinates-tagged tweets among geotagged ones range from 15\% to 30\%. However, certain countries like Turkey and Indonesia have extraordinary high percentages of coordinates-tags among geotagged tweets, which implies that it is easier to get precise location information for tweets from these countries. 

\begin{table*}[!th]
\centering
\caption{The geotagging distributions for users with different user sources (Top 10).}
\label{user_source}
\begin{tabular}{|l|r|r|r|}
\hline
User source              & Non-geotagged & Place-tagged  & Coordinates-tagged \\ \hline
Twitter for iPhone  & 5,741,431 (70.80\%) & 1,242,823 (15.33\%) & 1,125,263 (13.88\%)      \\ \hline
Twitter for Android & 4,869,846 (76.08\%) & 670,206 (10.47\%)   & 860,953 (13.45\%)        \\ \hline
Twitter Web Client  & 1,497,191 (77.99\%) & 214,419 (11.17\%)   & 208,006 (10.84\%)        \\ \hline
Facebook            & 289,930 (74.43\%)   & 29,737 (7.63\%)     & 69,869 (17.94\%)         \\ \hline
twittbot.net        & 301,390 (99.18\%)   & 1,076 (0.35\%)      & 1,408 (0.46\%)           \\ \hline
Twitter for iPad    & 210,894 (82.99\%)   & 19,759 (7.78\%)     & 23,482 (9.24\%)          \\ \hline
TweetDeck           & 215,006 (85.36\%)   & 16,709 (6.63\%)     & 20,163 (8.01\%)          \\ \hline
Twitter Lite        & 233,175 (93.24\%)   & 9,635 (3.85\%)      & 7,262 (2.90\%)           \\ \hline
Google              & 104,839 (86.39\%)   & 5,906 (4.87\%)      & 10,608 (8.74\%)          \\ \hline
Instagram           & 58,931 (55.44\%)    & 978 (0.92\%)        & 46,391 (43.64\%)         \\ \hline
\end{tabular}

\end{table*}

\subsection{Tweet Language}
Tweet language is an important proxy to understand locations of Twitter users. There are 65 languages detected by a BCP 47 language identifier\footnote{https://tools.ietf.org/html/bcp47} and those undetermined is denoted by "und". We present the geotagging distributions for different languages in Table \ref{tweet_lang}. Tweeters who use Japanese and Korean are the two most conservative groups of using geotags. Tweets written in Portuguese, Turkish, and Indonesian are the most likely to have geotags. For Turkish and Indonesian tweets, they have a much higher chance to be associated with precise coordinates. The statistical correlation is significant ($\chi^2$=331941651.52, 28df, p$<$0.001).

\section{User-Level Analysis}

\subsection{User Source}
We first select the most frequently used tweet source as the major source for each user, then divide users into categories based on their major tweet source. Table \ref{user_source} presents the geotagging distributions for users with different major sources. Again, the user-level geotagging distribution per source shows that iPhone users (29.2\%) are more likely to have geotags than Android users (23.92\%). However, even though the number of geotagged iPhone users is larger, the chance of they are coordinates-tagged is almost the same with Android users. Users who tweet from Instagram are the group with the highest chance of being coordinates-tagged (43.64\%), which is also consistent with the observation from tweet-level analysis.

\begin{figure}[!h]
    \centering
    \includegraphics[width=0.5\textwidth]{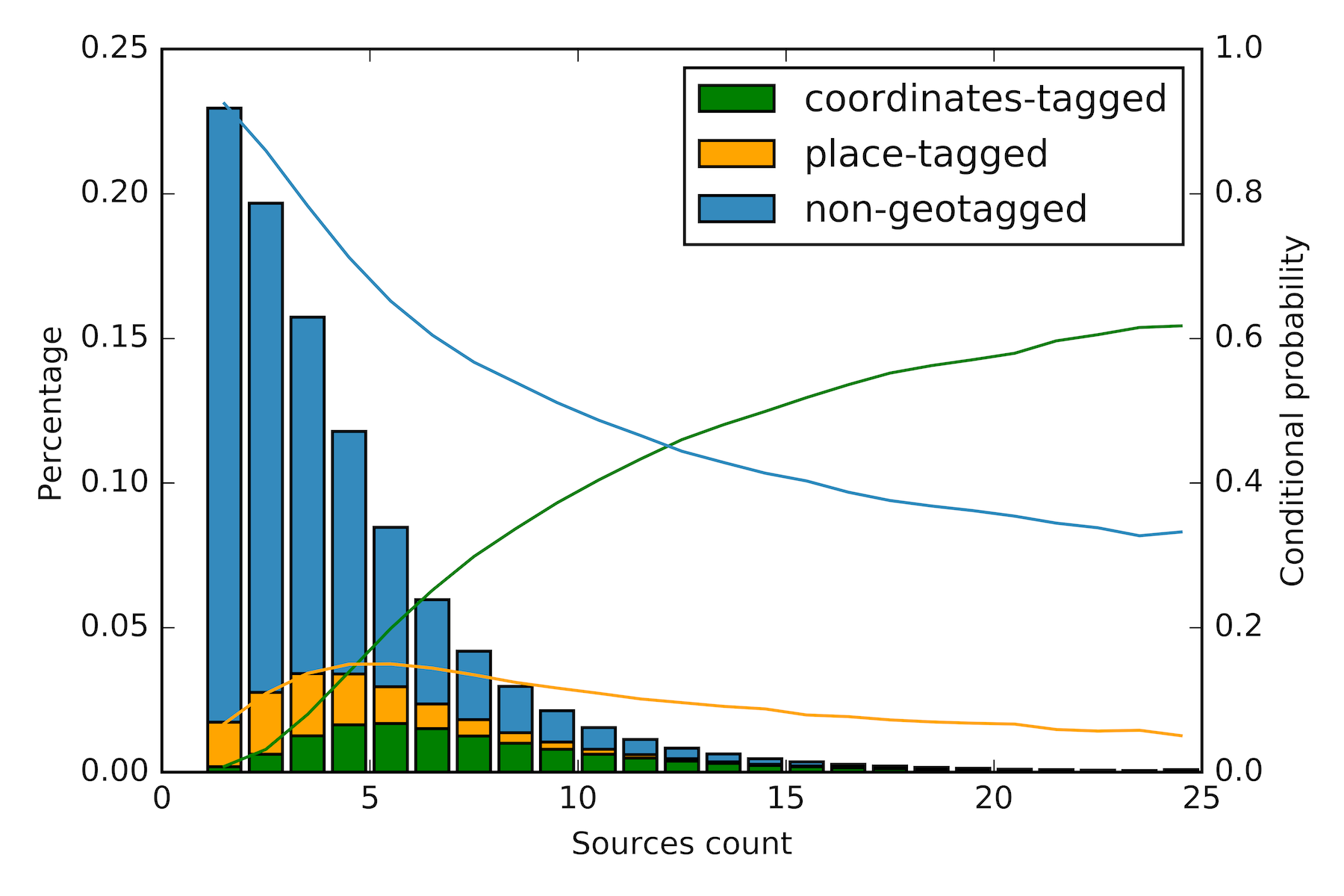}
    \caption{Left axis is associated with the bar chart of the distribution of users in terms of number of sources. Right axis is associated with the lines which represent the probabilities of place-tagging and coordinates-tagging given certain sources count. }
    \label{source_count}
\end{figure}

The bar chart in Figure \ref{source_count} shows the distribution of users in terms of the total number of sources they have used. The trending lines in this figure represent the probabilities of place-tagging and coordinates-tagging with varying sources count. As shown in Figure \ref{source_count}, the more sources a user have used before, the more likely he/she is geotagged and coordinates-tagged. 

\subsection{User Language}

Same as major source, we use the most frequently used tweet language as each user's major language. For the user major language, the differences among languages become even larger. As shown in Table \ref{user_lang}, users who use Korean are the most conservative group using geotags. Only 2.94\% of them are geotagged and 1.19\% are coordinates-tagged. On the contrary, for users speaking in Portuguese or Indonesian, we can get geotags for more than 35\% of them. More surprisingly, one third of users who use Indonesian have precise geo-coordinates which is much larger than the average percentage. The prevalence of geotags in users' timelines shows that it is possible to find location information for a large portion of users, especially for those speaking in languages like Indonesian or Portuguese.

\begin{table}[!h]
\centering
\caption{Geotagging distributions for different user languages (top 10).}
\label{user_lang}
\resizebox{0.5\textwidth}{!}{
\begin{tabular}{|l|r|r|r|}
\hline
User lang. & non-geotagged                                                 & place-tagged                                                  & coordinates-tagged                                            \\ \hline
English    & \begin{tabular}[c]{@{}r@{}}5,232,717\\ (69.77\%)\end{tabular} & \begin{tabular}[c]{@{}r@{}}1,157,694\\ (15.44\%)\end{tabular} & \begin{tabular}[c]{@{}r@{}}1,109,592\\ (14.79\%)\end{tabular} \\ \hline
Japanese   & \begin{tabular}[c]{@{}r@{}}3,539,998\\ (90.22\%)\end{tabular} & \begin{tabular}[c]{@{}r@{}}204,787\\ (5.22\%)\end{tabular}    & \begin{tabular}[c]{@{}r@{}}178,761\\ (4.56\%)\end{tabular}    \\ \hline
Spanish    & \begin{tabular}[c]{@{}r@{}}1,521,361\\ (66.63\%)\end{tabular} & \begin{tabular}[c]{@{}r@{}}278,065\\ (12.18\%)\end{tabular}   & \begin{tabular}[c]{@{}r@{}}483,997\\ (21.20\%)\end{tabular}   \\ \hline
Arabic     & \begin{tabular}[c]{@{}r@{}}1,326,259\\ (87.08\%)\end{tabular} & \begin{tabular}[c]{@{}r@{}}108,505\\ (7.12\%)\end{tabular}    & \begin{tabular}[c]{@{}r@{}}88,258\\ (5.79\%)\end{tabular}     \\ \hline
Portuguese & \begin{tabular}[c]{@{}r@{}}806,927\\ (64.12\%)\end{tabular}   & \begin{tabular}[c]{@{}r@{}}218,648\\ (17.37\%)\end{tabular}   & \begin{tabular}[c]{@{}r@{}}232,904\\ (18.51\%)\end{tabular}   \\ \hline
Korean     & \begin{tabular}[c]{@{}r@{}}538,095\\ (97.06\%)\end{tabular}   & \begin{tabular}[c]{@{}r@{}}9,702\\ (1.75\%)\end{tabular}      & \begin{tabular}[c]{@{}r@{}}6,584\\ (1.19\%)\end{tabular}      \\ \hline
Turkish    & \begin{tabular}[c]{@{}r@{}}311,746\\ (66.17\%)\end{tabular}   & \begin{tabular}[c]{@{}r@{}}49,714\\ (10.55\%)\end{tabular}    & \begin{tabular}[c]{@{}r@{}}109,671\\ (23.28\%)\end{tabular}   \\ \hline
French     & \begin{tabular}[c]{@{}r@{}}345,166\\ (75.65\%)\end{tabular}   & \begin{tabular}[c]{@{}r@{}}59,595\\ (13.06\%)\end{tabular}    & \begin{tabular}[c]{@{}r@{}}51,531\\ (11.29\%)\end{tabular}    \\ \hline
Thai       & \begin{tabular}[c]{@{}r@{}}301,935\\ (78.89\%)\end{tabular}   & \begin{tabular}[c]{@{}r@{}}33,497\\ (8.75\%)\end{tabular}     & \begin{tabular}[c]{@{}r@{}}47,283\\ (12.35\%)\end{tabular}    \\ \hline
Indonesian & \begin{tabular}[c]{@{}r@{}}217,667\\ (57.53\%)\end{tabular}   & \begin{tabular}[c]{@{}r@{}}34,889\\ (9.22\%)\end{tabular}     & \begin{tabular}[c]{@{}r@{}}125,771\\ (33.24\%)\end{tabular}   \\ \hline
\end{tabular}
}
\end{table}

\subsection{User self-reported location}

In user's profile, there is an option for users to report their home locations. This is a text field that users can fill whatever they want, so it could be empty, or with unrelated text like ``in your heart'', or with meaningful location information like ``Pittsburgh, PA''. In this section, we examine whether such reporting behavior is correlated with the geotagging behavior.

We utilize an public accessible gazetteer called Geonames\footnote{http://www.geonames.org} to map user self-reported profile locations into meaningful geographical places. Among these 20 million users, 38.6\% of them do not provide any location information in this field, and 41.2\% provide recognizable locations. 

As shown in Table \ref{profile_location}, users with empty profile locations are also the group with the lowest probability to share their real-time locations. In the meanwhile, users with recognizable profile locations are also very likely to use geotags and even share precise coordinates. For these non-geotagged users, only 36.47\% of them report recognizable locations, while the percentages are 55.82\% for geotagged users and even 58.21\% for coordinates-tagged users. 

Previous work has shown that profile location is the most important feature for user location prediction \cite{han2014text, han2013stacking}

\begin{table}[]
\centering
\caption{Geotagging distribution for users with different types of profile locations.}
\label{profile_location}
\resizebox{0.5\textwidth}{!}{
\begin{tabular}{|l|r|r|r|}
\hline
Profile location & Non-geotagged                                                 & Place-tagged                                                  & Coordinates-tagged                                            \\ \hline
Empty            & \begin{tabular}[c]{@{}r@{}}6,489,046\\ (84.09\%)\end{tabular} & \begin{tabular}[c]{@{}r@{}}625,701\\ (8.11\%)\end{tabular}    & \begin{tabular}[c]{@{}r@{}}602,036\\ (7.80\%)\end{tabular}    \\ \hline
Unrecognized     & \begin{tabular}[c]{@{}r@{}}3,111,036\\ (77.09\%)\end{tabular} & \begin{tabular}[c]{@{}r@{}}446,833\\ (11.07\%)\end{tabular}   & \begin{tabular}[c]{@{}r@{}}477,919\\ (11.84\%)\end{tabular}   \\ \hline
Recognized       & \begin{tabular}[c]{@{}r@{}}5,512,198\\ (66.96\%)\end{tabular} & \begin{tabular}[c]{@{}r@{}}1,215,208\\ (14.76\%)\end{tabular} & \begin{tabular}[c]{@{}r@{}}1,504,087\\ (18.27\%)\end{tabular} \\ \hline
\end{tabular}
}
\end{table}

\begin{figure}[h]
    \centering
    \includegraphics[width=0.5\textwidth]{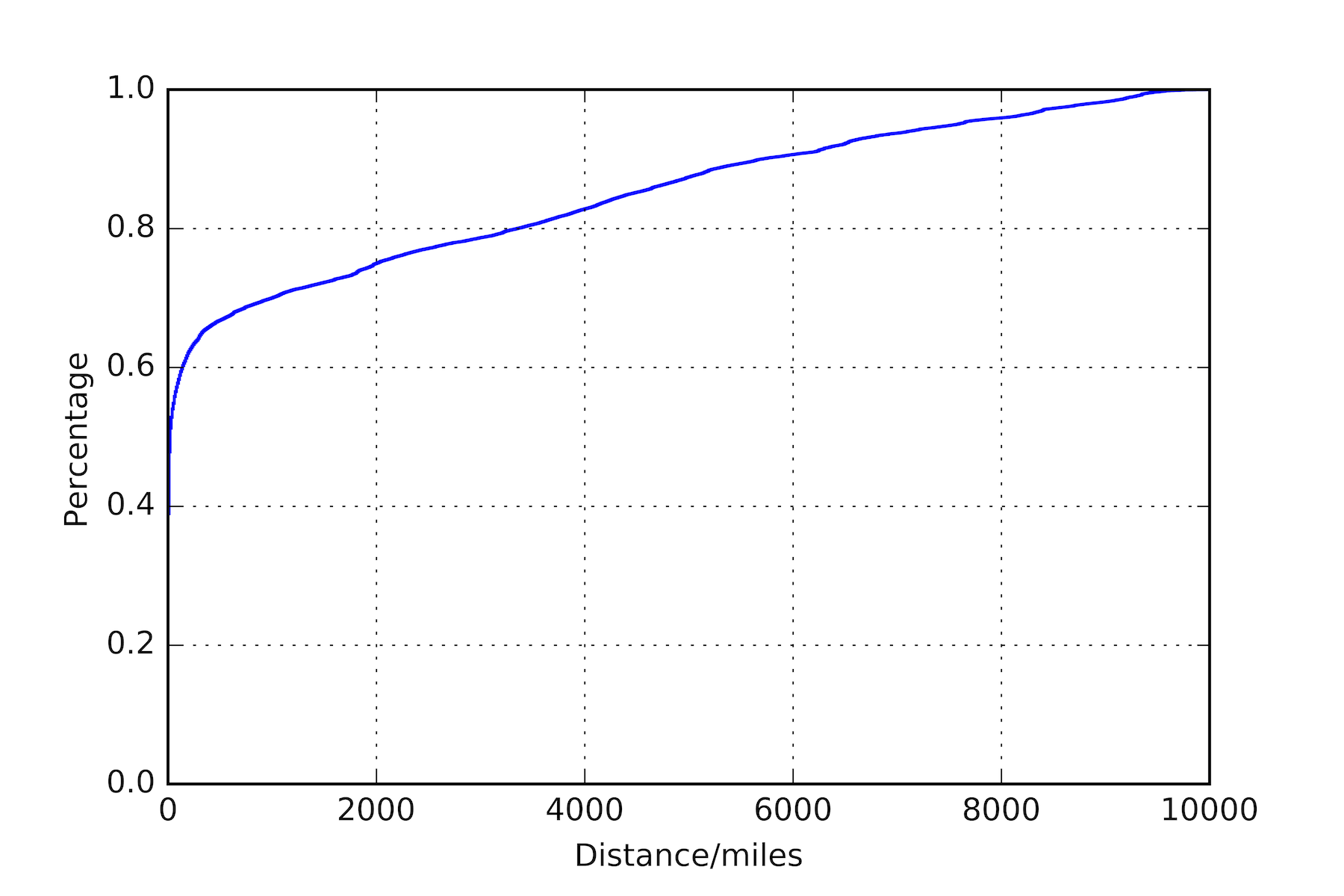}
    \caption{The cumulative distribution for distance between recognized location and the closest geotags.}
    \label{distance}
\end{figure}

Among these users with recognized profile locations, we extract 2,719,295 users that have accurate geotags (coordinates-tags or at least city-level place-tags). For each user, we calculate the distance between the recognized location and the closest geotags. We present the cumulative distribution of these distances in Figure \ref{distance}. As shown in this figure, 38.93\% of the distances are within 10 miles and 58.86\% are within 100 miles, which means that we can easily find home locations for a majority of self-reported users.

\section{Graph-level analysis}
In this section, we would like to explore whether the geotagging preferences of one user's followees or followers would have a correlation with this user's geotagging behavior. 

\begin{figure}[h]
    \centering
    \includegraphics[width=0.48\textwidth]{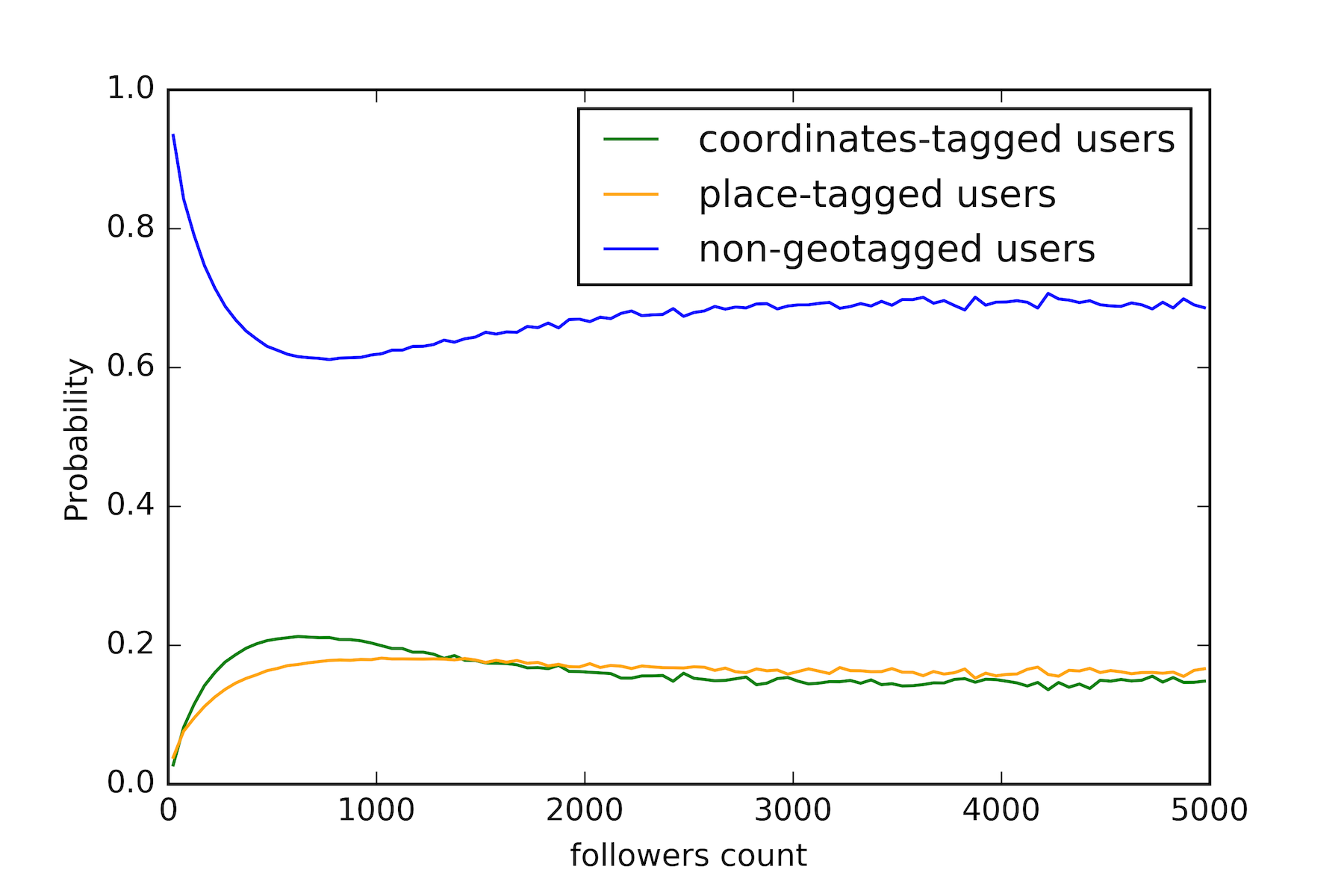}
    \includegraphics[width=0.48\textwidth]{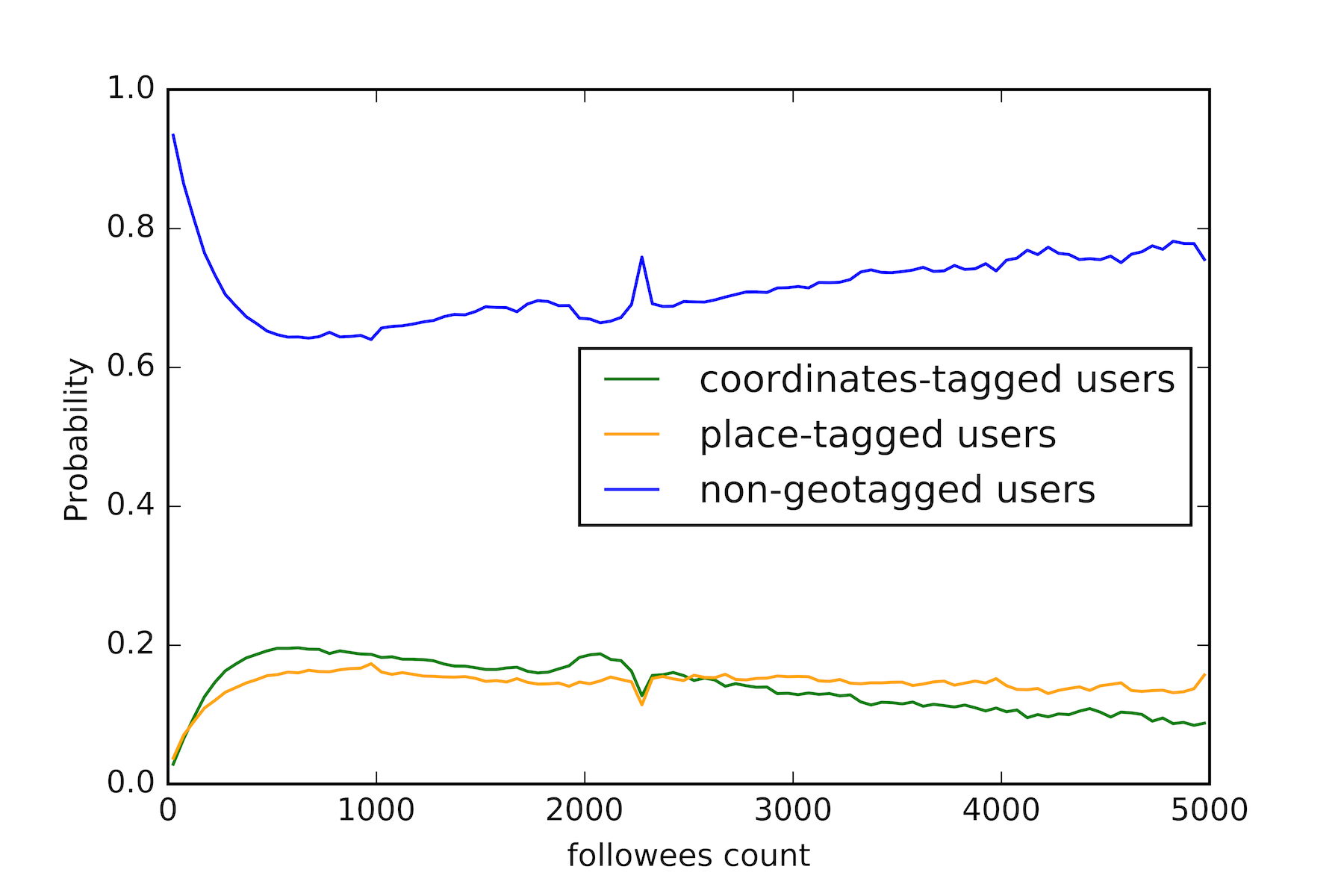}
    \caption{The percentages of different types of users with various number of followers/followees.}
    \label{follower}
\end{figure}

\begin{figure*}[!h]
\centering
    \begin{subfigure}[t!]{.48\textwidth}
        \centering
    \includegraphics[width=\linewidth]{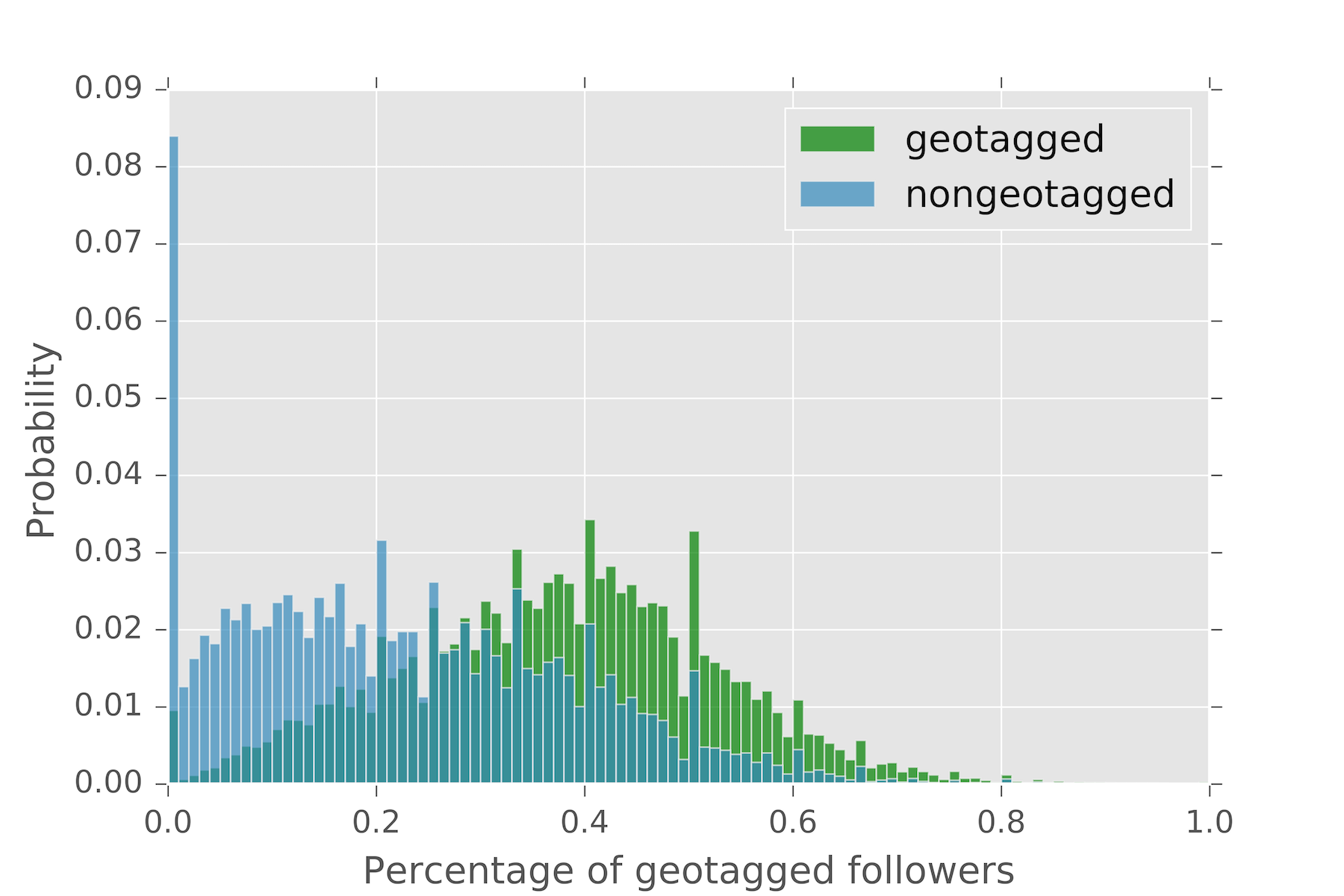}
     \caption{The distributions of the percentages of geotagged followers for geotagged and non-geotagged users.}
    \label{dist_follower_a}
    \end{subfigure}
        \begin{subfigure}[t!]{.48\textwidth}
            \centering
    \includegraphics[width=\linewidth]{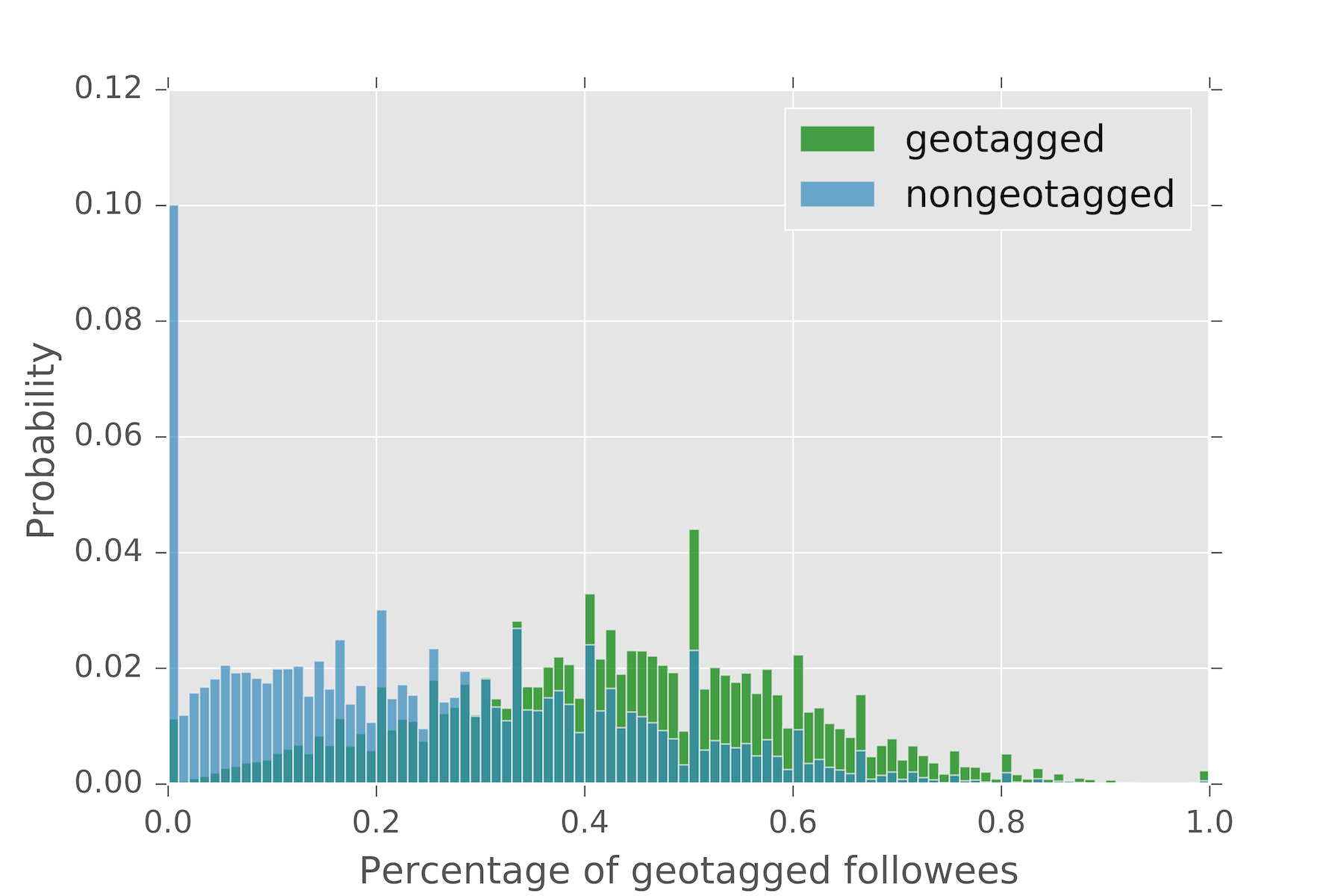}
     \caption{The distributions of the percentages of geotagged followees for geotagged and non-geotagged users.}
    \label{dist_follower_b}
        \end{subfigure}
        
    \begin{subfigure}[t!]{.48\textwidth}
                \centering
    \includegraphics[width=\linewidth]{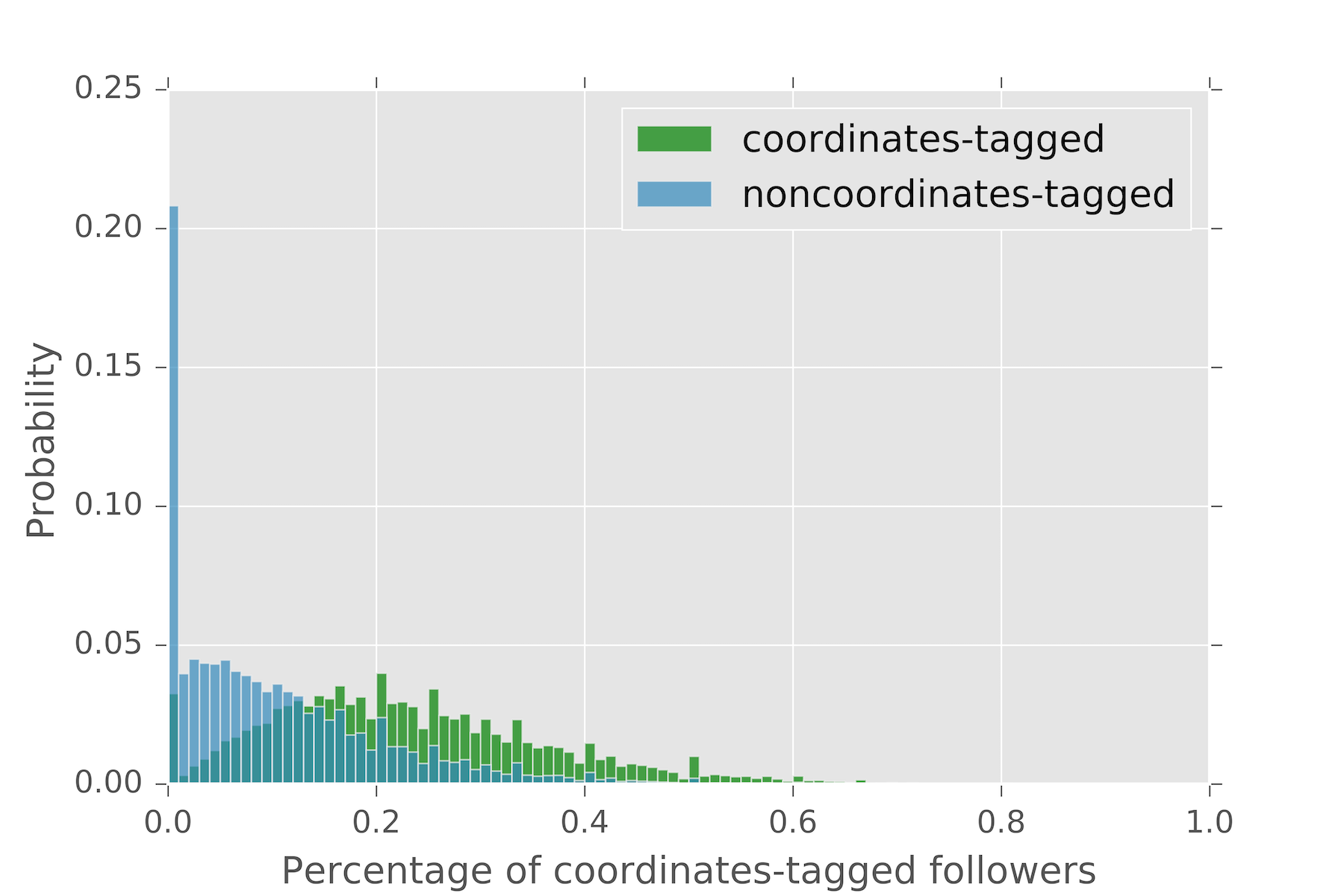}
     \caption{The distributions of the percentages of coordinates-tagged followers for coordinates-tagged and noncoordinates-tagged users.}
    \label{dist_follower_c}
    \end{subfigure}
        \begin{subfigure}[t!]{.48\textwidth}
          \centering
    \includegraphics[width=\linewidth]{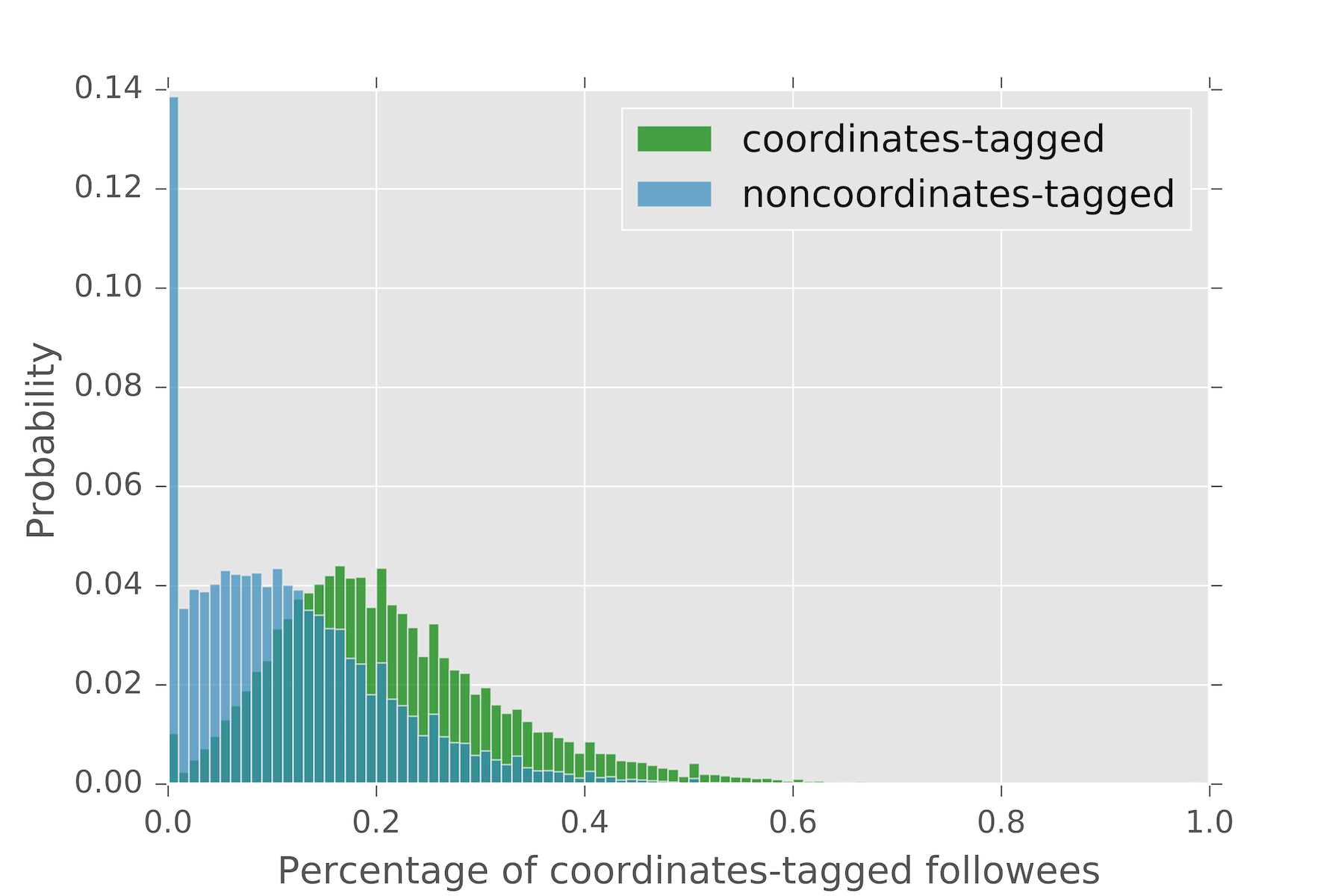}
     \caption{The distributions of the percentages of coordinates-tagged followees for coordinates-tagged and noncoordinates-tagged users.}
    \label{dist_follower_d}
    \end{subfigure}
    \caption{}
    \label{dist_follower}
\end{figure*}

We first calculate the percentages of non-geotagged, place-tagged, and coordinates-tagged users with various number of followers and followees. As shown in Figure \ref{follower}, the chance of one user sending real-time location information first increases rapidly with increasing number of followers and followees, then drops slowly after reaching certain thresholds. The percentage of coordinates-tagged users drops faster than place-tagged users. All imply that users with larger number of followers and followees tend to be more conservative about their location privacy.

To examine the homophily effect \cite{mcpherson2001birds}, we further compare the percentages of geotagged/coordinates-tagged followers and followees for different types of users. In this analysis, we only include users with at least five followers/followees.

In Figure \ref{dist_follower_a}, we plot the distributions of the percentages of geotagged followers for geotagged and non-geotagged users. As shown in Figure \ref{dist_follower_a}, geotagged users tend to have more geotagged followers. On average, 37.41\% of the followers of geotagged users are geotagged while only 22.86\% of the followers of non-geotagged users are geotagged. We further use Kolmogorov-Smirnov (KS) statistic to measure the difference of these two distributions. The KS test for these two distributions is 0.368 (p$<$0.001), which shows significant difference. Similarly, the average percentage of geotagged followees of geotagged users is 39.49\% while it is only 27.05\% for non-geotagged users. We also compared the distributions of the percentages of geotagged followees for geotagged and non-geotagged users. The KS test is 0.255 (p$<$0.001) for these two distributions.

Same type of comparisons are performed on coordinates-tagged users and noncoordinates-tagged users. The average percentage of coordinates-tagged followers of coordinates-tagged users is 23.09\% while it is only 10.54\% for noncoordinates-tagged users. The average percentages of coordinates-tagged followees are 21.34\% and 11.50\% for coordinates-tagged and noncoordinates-tagged users resepectively.

The KS test comparing two distributions of percentages of coordinates-tagged followers is 0.431 (p$<$0.001). The comparison is 0.395 (p$<$0.001) for two distributions of percentages of coordinates-tagged followees. All show significant differences among coordinates-tagged and noncoordinates-tagged users.

In addition, we compare the probability of an ego node being geotagged given at least of one alter is geotagged with the probability of an ego being geotagged given no alter is geotagged. In Table \ref{cond_geo}, we present these two conditional probabilities and show that given at least one alter is geotagged will increase an ego's chance of being geotagged by at least 6 times. Same thing also happens for coordinates-tagging as shown in Table \ref{cond_coo}.

\begin{table}[!h]
\caption{The comparison of conditional geotagging probabilities. A friend is the user who has a bidirectional link to the ego.}
\label{cond_geo}
\begin{tabular}{|l|c|c|}

\hline
      Alter            & \begin{tabular}[c]{@{}l@{}}P(Ego is geotagged\\  $|$ at least one alter \\ is geotagged)\end{tabular} & \begin{tabular}[c]{@{}l@{}}P(Ego is geotagged\\ $|$ no alter is geotagged)\end{tabular} \\ \hline
follower & 28.70\%                                                                                             & 4.17\%                                                                                \\ \hline
followee & 26.06\%                                                                                             & 1.76\%                                                                                 \\ \hline
friend   & 30.60\%                                                                                             & 4.40\%                                                                                 \\ \hline
\end{tabular}

\end{table}

\begin{table}[!h]
\caption{The comparison of conditional coordinates-tagging probabilities. A friend is the user who has a bidirectional link to the ego.}
\label{cond_coo}
\resizebox{0.5\textwidth}{!}{
\begin{tabular}{|l|c|c|}

\hline
         Alter         & \begin{tabular}[c]{@{}l@{}}P(Ego is coordinates-\\ tagged $|$ at least one alter \\ is coordinates-tagged)\end{tabular} & \begin{tabular}[c]{@{}l@{}}P(Ego is coordinates-\\ tagged $|$ no alter is \\ coordinates-tagged)\end{tabular} \\ \hline
follower & 16.74\%                                                                                                               & 2.75\%                                                                                                      \\ \hline
followee & 14.82\%                                                                                                               & 1.22\%                                                                                                      \\ \hline
friend   & 18.01\%                                                                                                               & 2.97\%                                                                                                      \\ \hline
\end{tabular}
}
\end{table}

\section{Discussion}
This study shows that users vary systematically in their willingness to share their real-time locations. Those who speak Turkish, Portuguese or Indonesian are more likely to share geotagged content; whereas, Korean and Japanese language users are unlikely to share such information. This implies that geotagged content, may not be representative of public opinion in the corresponding region. For example, if only using geotagged data one finds more discussion in Turkish about climate change than in Japanese, that does not mean that the Japanese are not concerned.  It might simply reflects the fact that more Turkish people are willing to post geotagged tweets. On the other hand, knowing the likelihood that tweets are geotagged by language, means that researchers can weight tweets by that likelihood to better estimate the true underlying population opinion. This finding also suggests that research focusing on regional opinions using geotagged tweets, will be more comprehensive if done in countries or regions where the fraction of people speaking those languages where users are likely to geotag is higher. In other words, such research may be more accurate if done in Portugal or Indonesia or Turkey. Future research should take the local linguistic variance into account to get a good estimation.

The heterogeneity of Twitter users not only affects the regional analytics in Twitter, but also impacts the generalizability of location prediction systems. In our user-level analysis, we found that Twitter users' geotagging behavior and location reporting behavior are highly correlated. Users who report locations in their profiles are more likely to use geotags. Todate, most Twitter geolocation classifiers are built based on the features learned from geotagged users \cite{huang2017predicting,graham2014world}.  The research reported here suggests that such systems may be less useful than previously thought, because a disproportionate number of users that use geotags also report locations.  Further, there is no reason to believe that those who geotag are distributed in terms of locations the same as those who do not geotag. Research needs to examine whether there are fundamental user characteristics, such as psychological preferences, political sensitivities, or socialized behavior that are impacting the usage of geotags and so locations. 

Finally, we found a strong homophily effect in Twitter users' geotagging preferences. Specifically, geotagged users tend to connect to geotagged users while non-geotagged users tend to connect to non-geotagged users. Given at least one alter is geotagged will increase an ego's chance of being geotagged by at least 6 times. These findings also call into question the validity of graph-based location prediction systems. If non-geotagged users tend to cluster together, then it becomes harder to find non-geotagged users' location based on the information from their friends. It also suggests that geotagging may be a behavior that is prone to social influence \cite{marsden1993network}.

In conclusion, this paper looks at the user geotagging behavior from 20 million randomly sampled Twitter users' timelines. The answer to our three research questions is yes. There are three main insights that certain types of users are more willing to share their location information via geotagging, these less conservative users are also more likely to self-report their locations in profiles, and users with similar geotagging preferences tend to cluster together. This research provides an empirical basis that helps inform research on geotagged social media like regional user opinion analytics, spatial event detection, and location inference.

\section*{Acknowledgment}
We want to thank all the members in the Computational Analysis of Social and Organizational Systems at Carnegie Mellon University for their valuable discussion. This work is supported by the Office of Naval Research (ONR) N00014182106 and N0001418SB001. The views and conclusions are those of the authors and should not be interpreted as representing the official policies, either expressed or implied, of the ONR.

\bibliographystyle{IEEEtran}
\bibliography{./content/reference}

\end{document}